\documentclass[12pt,preprint]{emulateapj}

\begin{document}
\title{Feedback Effects of First Supernovae on the Neighboring Dark
       Matter Halos}
\author{Masaru Sakuma}
\affil{Center for Computational Sciences, University of
  Tsukuba, Tsukuba, Japan}
\email{yusaku@ccs.tsukuba.ac.jp}

\author{Hajime Susa}
\affil{Department of Physics, Konan University, Kobe, Japan}
\email{susa@konan-u.ac.jp}

 \begin{abstract}
  The first-generation
  stars in the $\Lambda$CDM universe are considered to have formed in
  dark halos with total masses in the range 
  $\sim 10^{5}-10^{7}\rm{M_\odot}$ at $z\sim 20-50$.  These stars
  expected to be very massive and in some cases they end their lives as
  the first supernovae (SNe).  We explore the problem of
  whether star formation in low mass dark halos ($\leqslant 10^7\rm{M_\odot}$)
  was triggered or suppressed by the SN feedback from neighboring
  star-forming halos.  We take into consideration mainly two effects by
  the SN shock: one is the evacuation of gas components from the halos
  and the other is the promotion of H$_2$ formation because of the
  enhanced ionization degree by shock heating.  Combining above effects,
  we find that the star formation activities in the neighboring dark
  matter halos ($M\leqslant 10^7\rm{M_\odot}$) are basically suppressed in
  case they are located close to the SN center, because of the gas
  evacuation effect.  The critical distance within which the gas is
  blown away falls within the range $\sim 0.3-1.5$kpc depending on the SN
  energy and the halo mass.  In addition, we find there is very little
  window in the parameter space where star formation activities in dark
  halos are induced or promoted by neighboring SN. 
 \end{abstract}


\section{INTRODUCTION}
Star formation in the early universe has played a critical role in
subsequent evolution of the universe.  First-generation stars may have
substantially contributed to the cosmic reionization and metal pollution
of the universe by their radiative/kinetic feedback effects.  These
events are important for the formation and evolution of protogalaxies. 
The studies of the initial collapse of primordial pre-galactic objects in
the $\Lambda$CDM universe have been done in both analytical \citep
{Tegmark97,NS99} and numerical method \citep{AANZ98,FC00,Yoshida03}.  
These objects are formed with masses of the order of 
$\sim 10^{5}-10^{7}\rm{M_{\odot}}$ at redshifts $z\sim 20 - 50$.
Theoretical studies suggested that in these low-mass halos, $\rm{H_{2}}$
molecules are formed up to the level of $\sim 10^{-4}$, because the
virial temperature and the central density of these halos 
are high enough to activate H$_2$
formation \citep {NS99} so that the cooling time becomes short enough.  
This small fraction of H$_2$ is sufficient to
cool the gas, which leads to the formation of first-generation stars in
these halos.  

Early studies on the formation of primordial stars have been done in almost
one-zone approximation \citep
{MST65,Hutch76,Carl81,PSS83,SPS86a,SPS86b,SUN96}. 
On the other hand, after 1990s, there have been a number
of numerical studies of the formation of primordial stars \citep 
{ON98,NU99,BCL99,BCL02,ABN00,ABN02,OP01,OP03,Gao05,Yoshida06,O'N07}.
These studies consistently suggested that first-generation stars are
very massive ($\sim 30 - 500\rm{M_\odot}$). Because of their extreme
mass scale, first-generation stars emit copious amount of ionizing
radiation, as well as a strong flux of $\rm{H_{2}}$-dissociating
Lyman-Werner (LW) band radiation. Therefore, the radiation from the
first stars dramatically influences their surroundings, heating and
ionizing the gas within a few kiloparsecs around the progenitor star. 

If the primeval star is very massive ($\gtrsim 100\rm{M_{\odot}}$), the
ionization front from the star will break out of parent halos up to
$10^{7}\rm{M_{\odot}}$ in mass.  Half of the baryons in the halo will be
swept up into a dense shell that grows to the virial radius of the halo
($\sim 100\ \rm{pc}$) by the end of the life of the star.  Therefore, the
gas density within the shell radius will drastically decrease to low uniform
densities of $0.1 - 1 \rm{cm^{-3}}$ prior to the supernova (SN)
explosion \citep {Wha04,Kitayama04,Alv06,AWB07,WA07}.

The dark halos are also affected by the radiation from the
first star in the neighboring halos.  The local radiative feedback
effect is sensitive to the distance from the source star, as well as 
the mass and the evolutionary stage of the target halo.
The ionizing radiation from the star totally or partially photoevaporates
them before the SN blast ever reaches them \citep
{O'shea05,Susa06,MBH06,AS07,Wha08a}. Therefore, these halos will be less
massive or be gone altogether by the time the SN shock strikes them. 
It is also should be noted that the LW band radiation from the
source star photodissociates H$_2$ molecules in these halos in case the
halos are located close enough to the source \citep{GB01,Susa07}.  

Some of massive primordial stars end their lives as energetic
SNe. \citet {Heger02} suggested that the progenitor star of 
$10 - 40\rm{M_{\odot}}$ dies in Type II SN blasts while those of 
$140 - 260\rm{M_{\odot}}$ dies in pair instability SN. On the other
hand, it is important to note that some POP III stars lying in between
the ranges indicated above may also explode as hypernovae \citep
{TUN07}.  The studies of first SN explosions have been performed in both
SPH \citep {BYH03,Greif07} and grid codes \citep {Kitayama05,Wha08b}.
In less massive halos ($M \la 10^7\rm{M_\odot}$), 
the ejecta first expands into a nearly uniform rarefied ionized medium,
 then interacts violently with the dense shell swept up in
the progenitor H {\scriptsize II} region. Part of the energy of the
blast wave is reflected into the center as a reverse shock and the rest
pushes the shell forward \citep {Kitayama05,Wha08b}.  
In massive parent halos ($\gtrsim 10^{7}\rm{M_{\odot}}$), however, the
SN remnant (SNR) will only expand $20 - 40\ \rm{pc}$ into the halo and then
recollapse \citep {Wha08b}, never reaching any nearby halos because the
{\it I}-front does not break out these parent halos prior to the SN
explosion.

The shock wave from an SN that break away from the host halo impact and
go through the neighboring halos those survived photoevaporation
process.  The shock wave may blow off the gas component;
however, it also promote the formation of $\rm{H_{2}}$ and HD molecules 
in the gas because of the enhanced electron abundance by the shock
heating \citep {SK87,KS92,Susa98,Ferrara98,NS99,OH02,UI00,JB06}. 
We also note that the relic H {\scriptsize II} region including the
photoevaporated halos recombines out of equilibrium, which is basically
same physical condition of the postshock enhancement of H$_2$/HD
formation. The SN shock can sweep up this H$_2$/HD and carry it into the
nearby halo. Therefore, these molecules in the fossil H {\scriptsize II}
region are important for the formation of secondary stars as well as
those would be formed in the SN shock by collisional ionization.  

Star formation could be triggered by SNRs in several ways. Primordial
SNRs may sweep up a dense shell of ambient gas or collide with the dense
shell swept up by an H {\scriptsize II} region, either of which 
becomes contaminated by metals in the ejecta, subsequently comes to be
dominated by its self-gravity.  As a result, the shell is expected to
fragment into smaller filaments/cores where populations of less massive
stars are formed \citep {Mac03,Salvaterra04,Machida05,Wha08b}.
On the other hand, in rather massive halos 
($\gtrsim 10^7\rm{M_{\odot}}$), the H {\scriptsize II} region generated
by the progenitor star is confined well inside the virial radius and the
gas is kept neutral. In this case, first SNRs expand in neutral halos,
heavily mix their interiors with heavy elements, and then recollapse 
without escaping the halo \citep {Wha08b}.  

The interaction of the first SN with neighboring halos in
the early universe has not been investigated in detail so far.
Recently, \citet{Greif07} performed a cosmological simulation of first
SN explosion.  They found that the shock from the SN can accelerate the
star formation process in neighboring rather massive halos in which
stars could be formed without feedback effects.  They also indicated
that if the SN explosion occurs in an H {\scriptsize II} region, the SNR
comes to pressure equilibrium at half of the radius of the relic
H {\scriptsize II} region because of its relatively large
temperatures. This limits the reach of the SN explosion to  $\sim
1.5\rm{kpc}$ \citep {Wha04,Kitayama04}.  In addition, \citet{CR08}
performed numerical simulation of SNRs ram-pressure stripping
cosmological halos. They focus on how heavy elements from the remnant
are mixed with the halo gas.  

However, the number of halos studied by numerical simulations is
restricted. More analytical criteria are needed in order to obtain a
systematic understanding of first SN feedback effects on neighboring
dark matter halos.  In this paper, we investigate feedback effects by
first SN onto the nearby halos in the early universe, which is
potentially important for the total star formation activities in the
early universe.  We use analytic arguments in order to obtain the
universal criteria for feedback effects by first SN on the neighboring
dark halos. The outline of this paper is as follows. The initial setup
and the description of SNR evolution are given in Section 2.  In Section
3 and 4, the SN feedback effects are listed and quantified. Section 5
and 6 are devoted to discussion and summary.

\section{MODEL}
We consider SN explosions in the first collapsed halos ($\la
10^{6}\rm{M_{\odot}}$). The gravitational potential of these halos are so
shallow that the ionizing radiation from the progenitor first stars can
sweep out the gas of the host halos
\citep{Wha04,Kitayama04,Alv06,AWB07,WA07}.  As a result, the subsequent
SNe can easily break away from the halo because of the decreased gas
density by photoionization prior to the explosion \citep{Kitayama05,Wha08b}.
In order to understand the nature of the SN shock expanding into an
essentially uniform and ionized intergalactic space, we assume spherical
symmetry and initially uniform ambient gas density of averaged
cosmological density, $\rho_0$.  Strictly speaking, this assumption is
not valid since the gas density of the halo still slightly higher than
$\rho_0$ even after the feedback of UV radiation.  This assumption
should have some effects to increase the energy of the SN shock, but we
use this for simplicity. 
The energy loss mechanism from the SNR is dominated by
bremsstrahlung for a first few years, then line emission, and then
inverse Compton scattering becomes important \citep{Kitayama05,Wha08b}. 
We take into account all of the cooling rate stated above as well as
$\rm{H_{2}}$ cooling at low temperature. We 
use the fitting formula for these rates from the compilation by
\citet{FK94} and \citet{GP98}, respectively. 

We consider the neighboring dark matter halos with total mass of
$10^5 \rm{M_\odot} \leqslant M_{\rm dh} \leqslant 10^7 \rm{M_\odot}$, 
whose baryonic fraction in mass is assumed to be the cosmic mean value, 
$\rm{\Omega_{\rm b}/\Omega_{\rm{M}}}$.  We have another free parameter
in our calculations, which is the distance from the SN center to the
target halo.  Additionally, we use two typical fixed redshift when
the SN explode ($z_{\rm SN} = 20$) and when the nearby halos
virialized ($z_{\rm vir} = 30$).  
We tested this issue with three
different SN explosion energy $E_{\rm SN} = 10^{51},10^{52}$, and
$10^{53} {\rm erg}$.  Throughout the paper, we work with the 
$\Lambda$CDM universe with $\rm{\Omega_{\rm{M}}} = 0.3$, 
$\rm{\Omega_{\rm{\Lambda}}} = 0.7$, 
$h = 0.7$, and $\Omega_{\rm{b}} = 0.05$.

\hspace{0.1in}

\subsection{Timescales}
We introduce three timescales $t_{\rm{s}}$, $t_{\rm{cool}}$, and
$t_{\rm{ff}}$, that characterize the important physical processes.  
They represent the sound crossing timescale of the gas component in
the neighboring halo, the cooling timescale, and the free-fall
timescale, respectively.  
The sound crossing timescale of the gas component in the neighboring
halo is described as
\begin{equation}
 t_{\rm s}(l) \equiv \frac{l}{c_{\rm s}}, \label{eq:t_s}
\end{equation}
where $l$ denotes the length that we are interested and $c_{\rm s}$ is
the sound speed of the gas, respectively. 
This timescale is also interpreted as the expansion timescale of the
shock heated gas if we substitute the size of the shock-heated region
for $l$ and use the sound speed of shocked gas as $c_{\rm s}$.  

The cooling timescale is defined as
\begin{equation}
 t_{\rm cool}(T) \equiv \frac{nkT}{(\gamma - 1) \Lambda (T,n,f_{\rm H_{2}})},
  \label{eq:t_cool}
\end{equation}
where $T$ and $n$ are the temperature and the number density, $k$ and
$\gamma$ denote the Boltzmann constant and the ratio of specific heats,
respectively.  $f_{\rm{H_{2}}}$ represents the fraction of $\rm{H_{2}}$,
and $\Lambda$ ($\rm{erg\ cm^{-3}\ s^{-1}}$) denotes the cooling rate of
gas.    
The free-fall timescale is written as
\begin{equation}
t_{\rm ff} \equiv \left(\frac{3\pi}{32G\rho_{\rm vir}}\right)^{1/2}.
\end{equation}

Here, $G$ is the gravitational constant and $\rho_{\rm vir}$ is the virial
density given by $\rho_{\rm vir} \equiv 18\pi^{2} \rho_{\rm cr}$, where 
$\rho_{\rm cr} \equiv 1.9 \times 10^{-29} h^{2}\ (1+z_{\rm vir})^{3}$ g 
$\rm{cm^{-3}}$.

\subsection{Evolution of SNR}
The time evolution of an SNR in intergalactic medium (IGM) is mainly
described by following four stages:

\begin{description}

 \item[(1)] {\it The free-expansion stage}. The free-expansion stage
	    lasts until the SN ejecta sweeps up roughly the same amount
	    of mass as their own in the surrounding medium.  In this
	    stage, the velocity of the SN ejecta decreases linearly with
	    radius \citep{Truelove99}.
 \item[(2)] {\it Sedov--Taylor adiabatic expansion stage}.  The
	    expansion of the shock front is well approximated by the
	    Sedov--Taylor solution. 
 \item[(3)] {\it Pressure-driven expansion stage}.  
	    After the postshock gas is cooled via the radiative
	    cooling, the SNR depart from an adiabatic
	    expansion.  Geometrically thin shell is formed just behind
	    the shock front. The expansion of the shocked shell is
	    driven by the high pressure of the hot cavity.
 \item[(4)] {\it Momentum-driven expansion stage}. In this stage, the
	    shocked shell expands conserving its momentum.

 \end{description}
In our calculation, we do not take into consideration the stage (1)
because the duration of this stage is very short and it hardly affects
the entire result.  We also do not consider the stage (4) in IGM since
the shock velocity at this stage is too small to activate H$_2$ molecule
formation or to evacuate the gas from the halo.  We consider the
scenario, in which the SNR gradually sweeps up mass and then collides
with the neighbor halo. In fact, the remnant first collides violently
with the dense shell swept up in the progenitor H {\scriptsize II} region,
but we leave this effect for future works, since introducing such effect
makes the analysis complicated.   

\subsubsection{\it The Sedov--Taylor adiabatic stage in IGM}
The expansion of the shock front in the Sedov--Taylor adiabatic stage is
described by the self-similar solution \citep{Sedov46,Taylor50}.  The
radius and the expansion velocity of the shock front is written as 
\begin{equation}
   R_{\rm S} = 1.15\ \left(\frac{E_{\rm SN}}{\rho_{0}}\right)
              ^{1/5} \ t^{2/5}, \label{eq:R_sedov}
\end{equation}
\begin{equation}
   v_{\rm S} = \frac{dR_{\rm S}}{dt} = 0.460\ 
                  \left(\frac{E_{\rm SN}}{\rho_{0}}\right)
                  ^{1/5}\ t^{-3/5}, \label{eq:v_Sedov}
\end{equation}
where $E_{\rm SN}$ and $\rho_{0}$ represent the SN explosion energy and
the ambient density, respectively.  The postshock temperature is derived
from the Sedov--Taylor solution and the
Rankine--Hugoniot relation as follows:
\begin{eqnarray}
 T_{\rm PS} &=& 0.423\ \left(\frac{\mu m_{\rm H}}{k}\right)\
  \frac{(\gamma - 1)}{(\gamma + 1)^{2}}\
  \left(\frac{E_{\rm{SN}}}{\rho_{0}}\right)^{2/5}\ t^{-6/5}\nonumber\\
   &=& 2.3 \times 10^{5}{\rm{K}} 
  \left(\frac{E_{\rm{SN}}}{10^{52}\rm{erg}}\right)
  \left(\frac{R_{\rm{s}}}{0.5\rm{kpc}}\right)^{-3},
\label{eq:T_sedov}
\end{eqnarray}
where $m_{\rm H}$ and $\mu$ denote the atomic mass unit and the mean
molecular weight, respectively. 
These solutions are based upon the assumption, that the gas is
adiabatic, however, the postshock gas forms a dense shell which is
cooled by the radiative cooling subsequently.  After the cooling
timescale of the shell becomes shorter than the expansion timescale of
the shell ($t_{\rm cool} < R_{\rm S}/v_{\rm S}$), 
the SNR move on to the pressure-driven expansion stage.

\subsubsection{\it The pressure-driven expansion stage in IGM}
The shock front expands by the high pressure in the hot cavity.  
We assume that the pressure inside the cavity decreases
adiabatically as
\begin{equation}
   P_{\rm ca} = P_{1}\ \left(\frac{R_{\rm S}}{R_{1}}\right)^{-3\gamma},
    \label{eq:P_ca}
\end{equation}
where $P_{\rm ca}$ denotes the pressure inside the cavity, and also 
$R_{1}$ and $P_{1}$ represent the radius of the shock front and the
postshock pressure at the beginning of the pressure-driven expansion
stage. Combining this equation with the equation of motion of the shell,
we have the shock radius/velocity as follows \citep{SI96}.
\begin{eqnarray}
   R_{\rm S}& =& 1.22\left( \frac{E_{\rm SN}}{\rho_{0}}\
           R_{1}^{\ 2}\right)^{1/7}\ t^{2/7}, \label{eq:R_pre} \\
   v_{\rm S}& =& 0.349\left( \frac{E_{\rm SN}}{\rho_{0}}\
           R_{1}^{\ 2}\right)^{1/7}\ t^{-5/7}. \label{eq:v_pre}
\end{eqnarray}

We also obtain the postshock temperature as follows:
\begin{eqnarray}
 T_{\rm PS}& =& 0.244\left(\frac{\mu m_{\rm H}}{k}\right)
  \frac{(\gamma - 1)}{(\gamma + 1)^{2}}
  \left( \frac{E_{\rm SN}}{\rho_{0}}\
   R_{1}^{\ 2}\right)^{2/7}\ t^{-10/7} \nonumber\\
 &=& 1.7 \times 10^{4}{\rm K}
  \left(\frac{E_{\rm SN}}{10^{52}\rm{erg}}\right)
  \left(\frac{R_{\rm{s}}}{1\rm{kpc}}\right)^{-5}.  
  \label{eq:T_pre}
\end{eqnarray}

\subsection{Metal cooling}
In the present paper, we do not take into account the cooling rate due
to heavy elements. In fact, the shock-heated gas in the nearby halos 
is expected to be polluted by metals ejected from the SN, although the
abundance of heavy elements in the shock-heated gas is hard to evaluate.
The ultra high resolution simulations would be indispensable to assess
the degree of metal mixing.  Aside from such difficulty, we can evaluate
the metal abundance assuming complete mixing between the SN ejecta and
the surrounding material.  Based upon such assumption, the mean
metallicity of the swept up mass by the SN shock is 
$Z/Z_{\odot} \sim 10^{-2.5}$ \citep{Salvaterra04, Greif07}. 
In addition, the metallicity of the high-{\it z} IGM observed by Ly$\alpha$
absorption systems is at a level of $Z/Z_{\odot} \sim 10^{-3}$ to
$10^{-2}$ \citep{Song01}.  Thus, $Z/Z_{\odot} \sim 10^{-3}$ to $10^{-2}$
could be a rough standard to assess the effects of metal cooling.
At such low metallicity, the radiative cooling rate is hardly affected
above $10^4$K \citep{BH89}, since it is dominated by H and He cooling. 
The cooling rate below $10^4$ K is basically proportional to the amount
of metals \citep{DM72,RCS76,RS77,SD93,BBC01}, however, again at such low
$Z$, the metal cooling rate is comparable to H$_2$ cooling, as long as 
$10^{3}\rm{K} \la T \la 10^{4}\rm{K}$ \citep{Susa04}.  Therefore,
radiative cooling by heavy elements in these halos do not play central
role as long as complete mixing is assumed. 

On the other hand, metals in the interior of expanding SNR will
mix with the shocked shell it sweeps up because of Rayleigh--Taylor and
Kelvin--Helmholtz instabilities.  This will enrich the shocked region to
much higher metallicities than $Z/Z_{\odot} \sim 10^{-3.5}$, radiatively
cool the shell much faster than any H$_2$ it has formed or swept up, and
then fragment clumps. It may be a clump that collides with the neighbor
halo, not an intact shell, but this is for future numerical studies to
determine. 

\begin{figure}[t]
 \begin{center}
  \includegraphics[width=\linewidth]{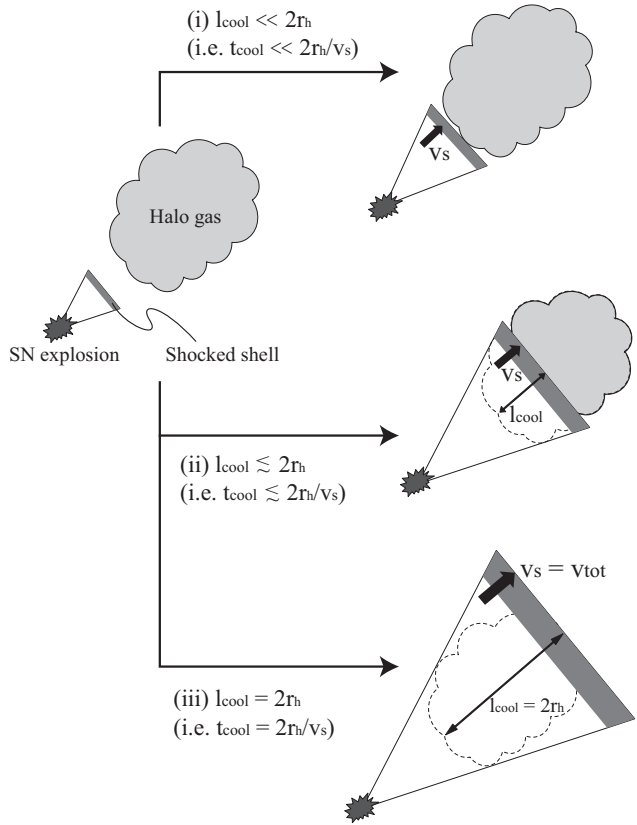}
  \caption{Three cases of gas evacuation by shock momentum are
  represented in this figure. The shock propagates at $v_{\rm S}$ before
  the shocked gas cools by radiative cooling. After that, the shock
  propagates under the momentum conservation. The intermediate case
  (case(ii)) indicates that the shock shifts to the momentum-driven
  expansion phase in the halo.}
  \label{fig:fig1}
 \end{center}
\end{figure}

\section{Effects of Supernova shock on the neighboring dark halos}
In this section, we assess the effect of the SN shock onto neighboring
dark halos by analytic arguments comparing timescales. 

\subsection{Gas evacuation by shock momentum}
In case a neighboring halo is located very close to the center of the SN
explosion, the gas in the halo would be evacuated by the shock momentum.
As a result, subsequent star formation in the halo should be inhibited.
Assuming the momentum conservation after the gas behind the shock cools, we
can roughly evaluate the velocity of the gas in the halo after the
impact of the shocked shell. 

In Figure \ref{fig:fig1}, we represent three cases of gas evacuation by
shock momentum.  First, we consider the simple momentum conserving case
(case(i) in Figure \ref{fig:fig1}),
where $l_{\rm cool} \ll 2r_{\rm h}$ (i.e., $t_{\rm cool} \ll
2r_{\rm h}/v_{\rm S}$) is
satisfied. Here, $l_{\rm cool}$ denotes the distance that the
shock sweeps the halo before it enters the momentum conserving phase, i.e.,
\begin{equation}
l_{\rm cool}\equiv \min\left(v_{\rm S}\ t_{\rm cool}(T_{\rm PS}),
			2r_{\rm h} \right).
\end{equation}

Here, $r_{\rm h}$ is the radius of the dark halo. 
The equation of the momentum conservation is given as follows: 
\begin{equation}
 m_{\rm S}v_{\rm S} = \left(m_{\rm S} + 
  \frac{\strut\displaystyle \rm{\Omega_{\rm b}}}
   {\strut\displaystyle \Omega_{\rm{M}}}{M}_{\rm dh}\right)v_{\rm tot},
\end{equation}
where $M_{\rm dh}$ is the mass of the dark halo and $m_{\rm S}$ is the
mass of the gas shell colliding with the halo, which is given as
\begin{equation}
  m_{\rm S} = \left(M_{\rm POPI\hspace{-.1em}I\hspace{-.1em}I}+\frac{4}{3}\pi
	      R_{\rm S}^{\ 3}\rho_{0}\right) \frac{\pi r_{\rm h}^{2}}
              {4\pi R_{\rm S}^{\ 2}}.
\end{equation}

Here, $M_{\rm POPI\hspace{-.1em}I\hspace{-.1em}I}$ is the mass of
progenitor POPIII star and we assume that the whole mass of this POPIII
star is released as ejecta. 
We assume $M_{\rm POPI\hspace{-.1em}I\hspace{-.1em}I} = 140\rm{M_\odot}$
in the present calculation.

On the other hand, in case of $t_{\rm cool} \la 2r_{\rm
h}/v_{\rm S}$ 
(i.e., $l_{\rm cool} \la 2r_{\rm h}$, case(ii) in Figure \ref{fig:fig1}),
the gas in the halo is cooled while the shock propagates in the halo by
radiative cooling, and it evolves into the momentum conserving phase. 
Since the gas pressure of the halo behind the shock front can keep
pushing the shock before it cools, $v_{\rm tot}$ cannot be estimated by
simple momentum conservation equation. It is difficult to assess this
effect analytically. 
Thus, we assume that the shock velocity is not affected by the halo while
the radiative cooling is still inefficient even after the shock enters
the halo. 
This assumption could be oversimplification, since even adiabatic shock
will slow down to some extent when it enters the dense region. In order
to quantify this effect, numerical study would be necessary, which is
beyond the scope of present study. 
Therefore, we have to keep in mind that the shock heating
effect is maximally taken into consideration in this model. 
%However, this effect would not be important since the shock velocity in
%the Sedov phase depends very weakly on the ambient matter
%density (see equation (\ref{eq:v_Sedov})). 
In this intermediate case ((ii) in Figure \ref{fig:fig1}),
we assume that the momentum conservation after shocked gas in the halo
is cooled: 
\begin{eqnarray} 
\left(m_{\rm S}+ \frac{\strut\displaystyle l_{\rm cool}}
   {\strut\displaystyle 2r_{\rm h}}\frac{\strut\displaystyle \rm{\Omega_{\rm b}}}
   {\strut\displaystyle \Omega_{\rm{M}}}M_{\rm dh}
    \right)v_{\rm S}
% &=& \left(m_{\rm S}+ \frac{\strut\displaystyle \rm{\Omega_{\rm b}}}
%   {\strut\displaystyle \Omega_{\rm{M}}}M_{\rm dh}
%   \frac{\strut\displaystyle l_{\rm cool}}
%   {\strut\displaystyle 2r_{\rm h}} + 
%   \frac{\strut\displaystyle \rm{\Omega_{\rm b}}}
%   {\strut\displaystyle \Omega_{\rm{M}}}M_{\rm dh}
%   \frac{\strut\displaystyle 2r_{\rm h} - l_{\rm cool}}
%   {\strut\displaystyle 2r_{\rm h}} \right)v_{\rm tot} \nonumber\\
 &=& \left(m_{\rm S}+ \frac{\strut\displaystyle \rm{\Omega_{\rm b}}}
   {\strut\displaystyle \Omega_{\rm{M}}}M_{\rm dh} \right)
   v_{\rm tot}. \label{eq:mon_cos}
\end{eqnarray}

In case the shock propagate through the entire halo within the
cooling time of the shocked gas, that is the case of 
$l_{\rm cool} = 2r_{\rm h}$, the shock velocity in the halo is as large
as $v_{\rm S}$ during the
shock propagation because of the inefficient cooling (case (iii) in
Figure \ref{fig:fig1}). 

The velocity of the shocked
gas shell, $v_{\rm tot}$, is obtained from the equation (\ref{eq:mon_cos}) as follows:
\begin{equation}
  v_{\rm tot} \equiv \frac{\ m_{\rm S}+\frac{\strut\displaystyle l_{\rm cool}}{\strut\displaystyle 2r_{\rm
   h}}
   \frac{\strut\displaystyle \rm{\Omega_{\rm b}}}
   {\strut\displaystyle \Omega_{\rm{M}}}M_{\rm dh}
    }
   { m_{\rm S} + \frac{\strut\displaystyle {\Omega_{\rm b}}}
   {\strut\displaystyle \Omega_{\rm{M}}}M_{\rm dh}\strut}\ v_{\rm S}.
		 \label{eq:v_tot}
\end{equation}
%Equation (11) represents two cases of gas evacuation by
%shock momentum.  If the shock propagate through the entire halo within
%the cooling time of the shocked gas, $t_{\rm cool}(T_{\rm PS})$, the
%shock velocity in the halo is as large as $v_{\rm S}$ because the
%shocked gas is not cooled during the shock propagation, i.e. the case of
%$l_{\rm cool} = 2 r_{\rm h}$. Otherwise the shock drastically weakened
%inside the halo by efficient cooling and the shock locally evolve into
%the momentum driven phase, i.e. the case of 
%$l_{\rm cool} = v_{\rm S}\ t_{\rm cool}(T_{\rm PS})$.  

This equation includes all of the three cases.
The cases of $l_{\rm cool} \ll 2r_{\rm h}/v_{\rm S}$ and 
$l_{\rm cool} = 2r_{\rm h}$ are the limits of efficient/inefficient cooling. 

\begin{figure}[t]
 \begin{center}
  \includegraphics[width=0.8\linewidth]{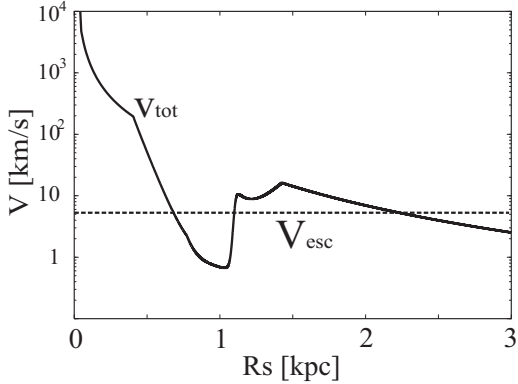}
  \caption{Velocity of the shocked gas ($v_{\rm tot}$) is plotted as
  a function of the distance from the SN center, $R_{\rm S}$, for 
  $E_{\rm SN} = 10^{52}\rm{erg}$ and 
  $M_{\rm dh} = 2\times 10^{5}\rm{M_{\odot}}$. The dotted line
  represents the escape velocity of the halo with 
  $M_{\rm dh} = 2\times 10^{5}\rm{M_{\odot}}$.}
  \label{fig:fig2}
 \end{center}
\end{figure}

The gas will be still bounded in the halo even after the shock arrival,
in case $v_{\rm tot}$ is smaller than the escape velocity $V_{\rm esc}$
of the halo.  On the other hand, if $v_{\rm tot} > V_{\rm esc}$, the gas
is blown away from the halo potential, in which subsequent star
formation in the halo is totally quenched.  Thus, we regard the condition
$v_{\rm tot} < V_{\rm esc}$ as a necessary condition for the triggered
star formation in the neighboring dark halos. 

In Figure \ref{fig:fig2}, $v_{\rm tot}$, as a function of the distance from
the SN center assuming $M_{\rm dh}=2\times 10^{5}\rm{M_{\odot}}$ and
$E_{\rm SN}=10^{52}\rm{erg}$.  
In this case, $v_{\rm tot}$ is smaller than $V_{\rm esc}$ in the ranges
$0.7{\rm{kpc}} \lesssim R_{\rm s} \lesssim 1.1\rm{kpc}$ and 
$R_{\rm s} \gtrsim 2.3\rm{kpc}$. 
Therefore, in case the neighboring halo is located at such distance, the
gas in the halo is not evacuated by the shock momentum.

\begin{figure}[t]
 \begin{center}
  \includegraphics[width=0.8\linewidth]{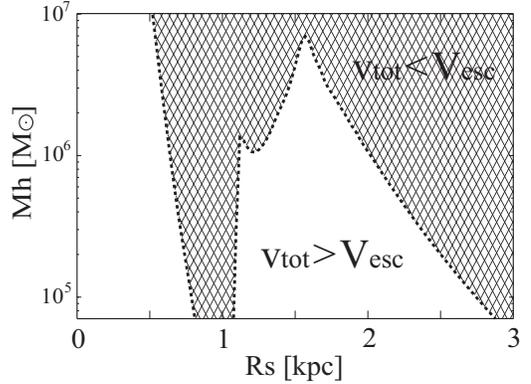}
 \caption{Condition of gas evacuation by the SN shock for
  $E_{\rm SN} = 10^{52}\rm erg$. The horizontal axis shows the distance
  between the nearby halo and SN center, while the vertical axis denotes
  the mass of the halo.  In the hatched area
  ($v_{\rm tot} < V_{\rm esc}$), the shocked gas is bounded inside the
  dark halo. The unhatched region labeled as  
  $v_{\rm tot} > V_{\rm esc}$, the gas will be evacuated by the shock
  from the SN.}
  \label{fig:fig3}
 \end{center}
\end{figure}

Figure \ref{fig:fig3} shows the gas evacuation from the halos for
$E_{\rm SN} = 10^{52}\rm{erg}$.  The vertical axis
denotes the mass of the dark matter halos ($M_{\rm dh}$), whereas the
horizontal axis represents the distance from the SN center.  
In the hatched area denoted as $v_{\rm tot} < V_{\rm esc}$, 
the gas in the dark halos are not lost by the shock momentum.
It is worth noting that the hatched area around $R_{\rm S}=1$kpc
corresponds to $T_{\rm PS}\sim 10^4$K,
where the Ly$\alpha$ cooling dominates the others.

\begin{figure*}[t]
 \epsscale{0.9}
 \begin{center}
  \plottwo{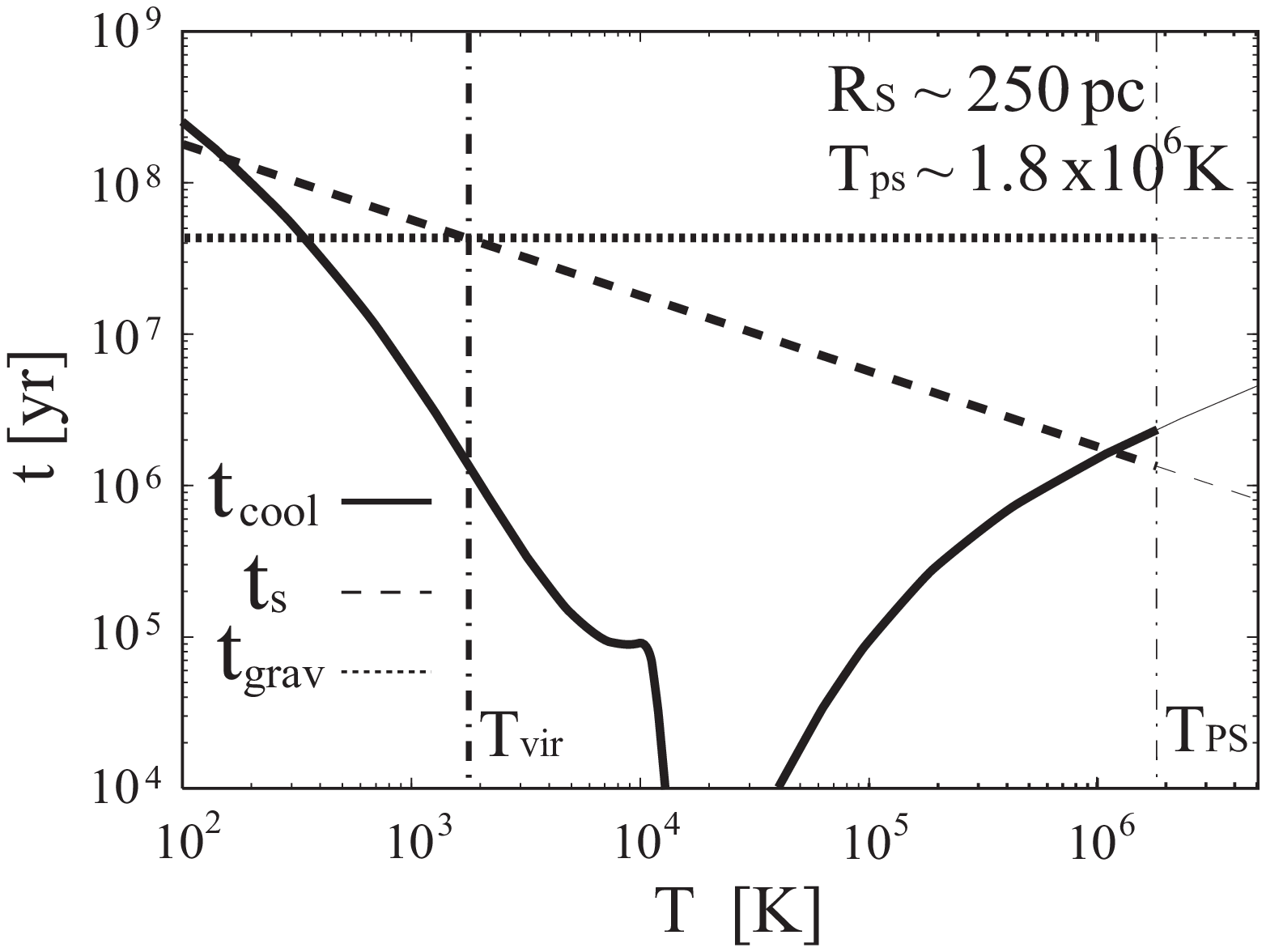}{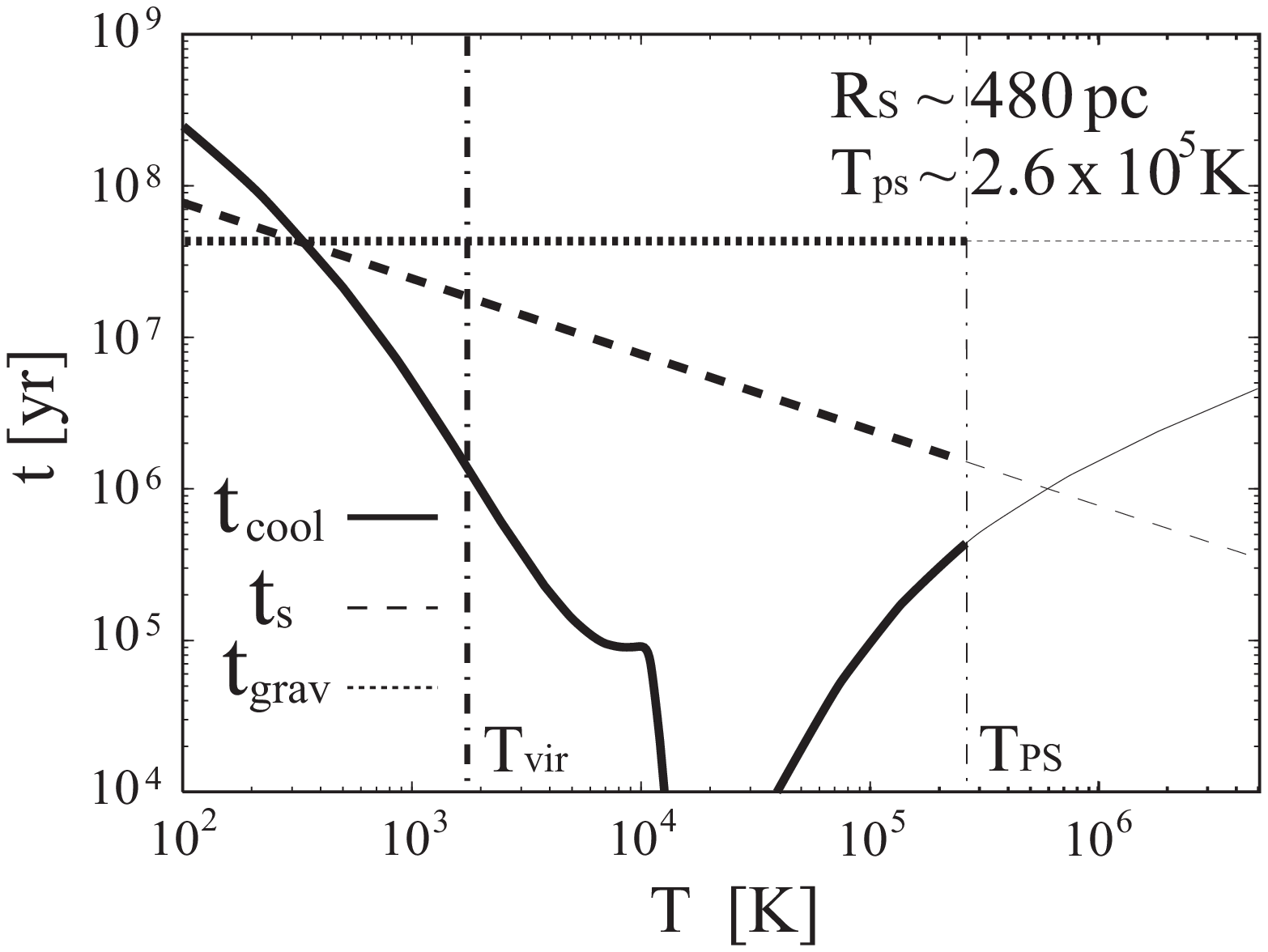}\
  \plottwo{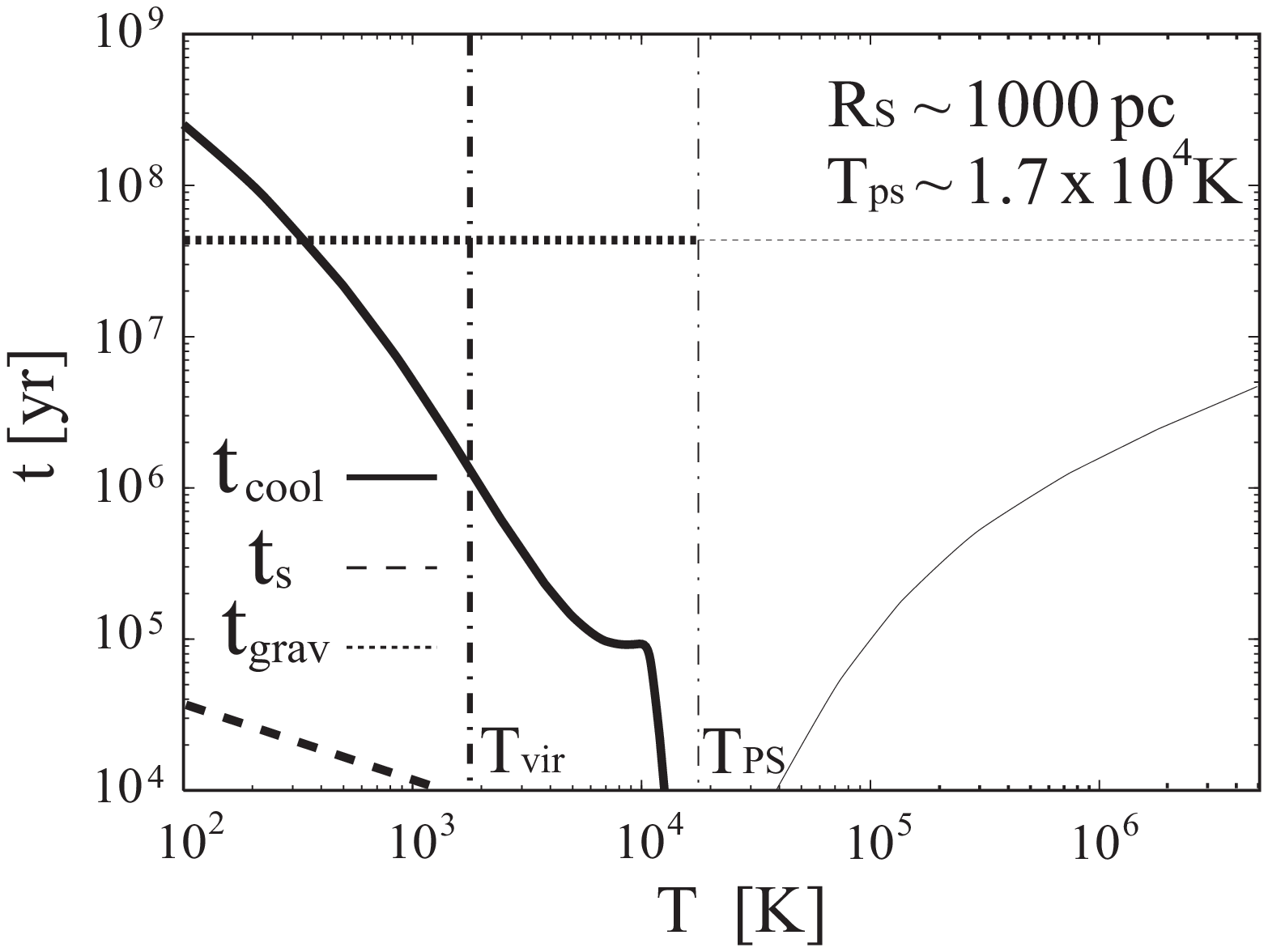}{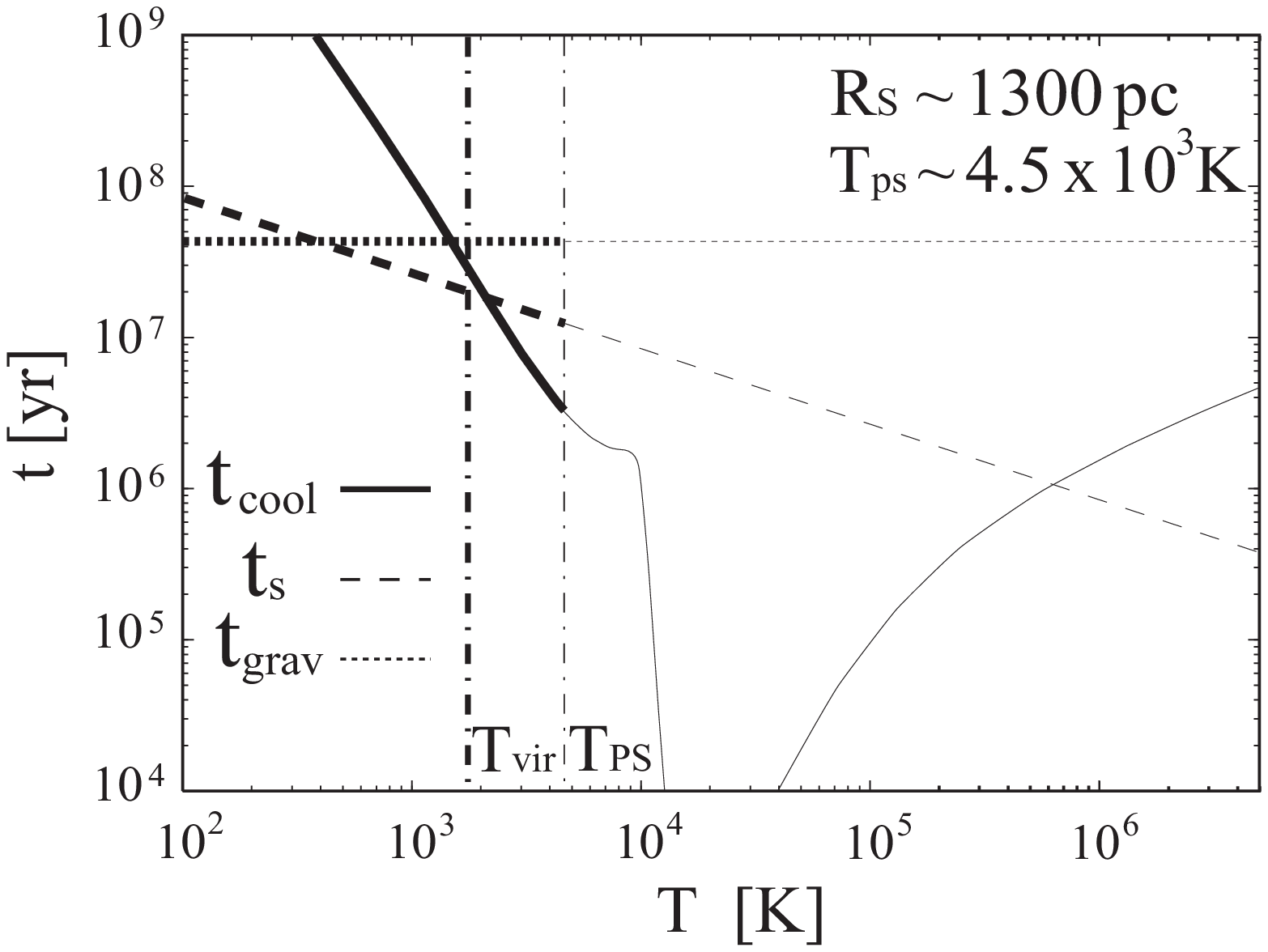}
  \caption{Evolution of timescales in the shock-heated gas is shown. 
  Four panels correspond to various distances
  $R_{\rm S}$ (upper-right corner), where the dark halos are located. 
  $t_{\rm cool},\ t_{\rm s},\ t_{\rm ff}$ represent the cooling
  timescale (solid), the sound crossing timescale of the shock-heated
  gas (dashed) , and the free-fall timescale (dotted), respectively. 
  The thick dot-dashed line indicates the virial temperature of the dark
  halo in the case of $M_{\rm vsh}=2\times 10^{5}\rm{M_{\odot}}$,
  whereas the thin dot-dashed line shows the postshock temperature 
  $T_{\rm PS}$.}
  \label{fig:fig4}
 \end{center}
\end{figure*}

\subsection{Cooling/collapse of the shocked gas}
The halos that satisfies the condition $v_{\rm tot} < V_{\rm esc}$
are able to survive the disruption by the SN shock momentum.  As a next
step, we consider whether those survived halos can collapse or not.  We
set the collapse criteria as the condition where the gas temperature is
decreased below $T_{\rm vir}$, before the gas cloud expands by the
thermal pressure, or bounces by adiabatic compression.  In other words,
if following conditions  
$$t_{\rm cool}(T) < t_{\rm s}(l_{\rm cool})~~~{\rm and}~~~ t_{\rm cool}(T) 
< t_{\rm ff}$$
are both satisfied until they cool below the initial virial temperature,
we regard that the gas in such halos can collapse to form stars.    
Here, we consider the sound crossing timescale, 
$t_{\rm s}(l_{\rm cool})=l_{\rm cool}/c_{\rm s}$, to be the expansion
timescale of the shocked region.  In general, the condition
$t_{\rm cool} < t_{\rm ff}$ is regarded as the collapse criterion of
gravitationally bound system \citep[e.g.,][]{RO77}. 
Once this condition is satisfied, the system starts to collapse. The
cooling timescale in low-density gas is basically inversely proportional
to the density, while the free-fall timescale proportional to the
inverse of square root of density. As a result, the ratio $t_{\rm cool}
/ t_{\rm ff}$ gets smaller as the collapse proceeds. That 's why the
condition is regarded as the collapse criterion. In case we consider the
primordial star formation, the cooling timescale in most of the final
run-away collapse phase is not proportional to the inverse of density,
however, the absolute value of the cooling timescale is shorter than the
other timescales outside the halo. Thus, the collapse of the cloud
continues following the track along which $t_{\rm cool}=t_{\rm ff}$ is
satisfied. 
In addition, we have to consider the adiabatic expansion of the shocked
gas, since the sound crossing timescale of the shocked gas could be very
short in the present case.  Thus, we need to add the inequality  
$t_{\rm cool} < t_{\rm s}(l_{\rm cool})$ to the condition of the
collapse criteria.  

In order to assess this condition, we have to follow the thermal
evolution of the shock heated gas in the dark halo.  
Each of the four panels in Figure \ref{fig:fig4}
shows the evolution of timescales after the shock heating for halos
with $M_{\rm dh}=2\times 10^{5}\rm{M_\odot}$ located at 
$R_{\rm s}=250{\rm pc},480{\rm pc},1000{\rm pc}$, and 
$1300{\rm pc}$. Thermal  energy of the SN explosion is assumed as 
$E_{\rm SN}=10^{52}{\rm erg}$. 
The horizontal axes denote the temperature of the gas on the way of
cooling, whereas the vertical axes show timescales.
In order to assess the cooling timescale for $T\la 10^4$K,
we have to take into account the H$_2$ cooling, which is proportional to
H$_2$ fraction. H$_2$ fraction could be obtained by solving
nonequilibrium chemical reaction equations, however, approximate values
are already obtained.  We use $f_{\rm H_2}=10^{-4}$ in case the
postshock temperature, $T_{\rm PS}$, is less than $10^4$K \citep{NS99}, 
whereas $f_{\rm{H_{2}}}=2\times 10^{-3}$ is employed for 
$T_{\rm PS} > 10^4$K \citep{SK87,Susa98,OH02}.  

If the halo with $M_{\rm dh}=2\times 10^5\rm{M_\odot}$ is located 
at $R_{\rm s}=250{\rm pc}$ (upper-left panel), the shocked temperature
is too high for the gas to remain the gas inside the halo potential.  
In fact, expansion timescale is already shorter than the cooling
timescale at $T=T_{\rm PS}$. Thus, the gas in this halo is lost
because of the shock heating.
On the other hand, in the cases of $R_{\rm S}=480{\rm pc}$ (upper-right
panel), $R_{\rm S}=1000{\rm pc}$ (lower-left panel), and 
$R_{\rm S}=1300{\rm pc}$ (lower-right panel), the gas in the halo can
start to cool because $t_{\rm cool}(T)$ is smaller than 
$t_{\rm s}(l_{\rm cool})$ at $T=T_{\rm PS}$.
Among these examples, $t_{\rm cool}(T)$
is shorter than $t_{\rm ff}$ and $t_{\rm s}(l_{\rm cool})$ as long as
$T>T_{\rm vir}$ is satisfied in the case $R_{\rm S}=480{\rm pc}$.  
Therefore, the gas can be cooled below
$T_{\rm vir}$ before it expands, and the cooling phase could be followed
by gravitational contraction.  
In contrast, in the cases of $R_{\rm S}=1000{\rm pc}$ and 
$R_{\rm S}=1300{\rm pc}$, the cooling process becomes inefficient before
$T=T_{\rm vir }$ is achieved.  In the case of $R_{\rm S}=1000 {\rm pc}$,
the postshock temperature exceeds $10^4$K. Thus, 
$f_{\rm H_2}=2\times 10^{-3}$ is achieved even for $T<10^4$K, whereas
$f_{\rm H_2}=10^{-4}$ for $R_{\rm S}=1300{\rm pc}$.  Therefore, the
cooling condition is not simply determined by the efficiency of H$_2$
formation, although it is necessary for the cooling below $10^4$K.  
 
The allowed region for cooling/collapse of the shocked gas is shown in
the next section.

\section{Possibility of triggered star formation in the neighboring dark
 halos}
\begin{figure}[t]
 \begin{center}
   \includegraphics[width=0.8\linewidth]{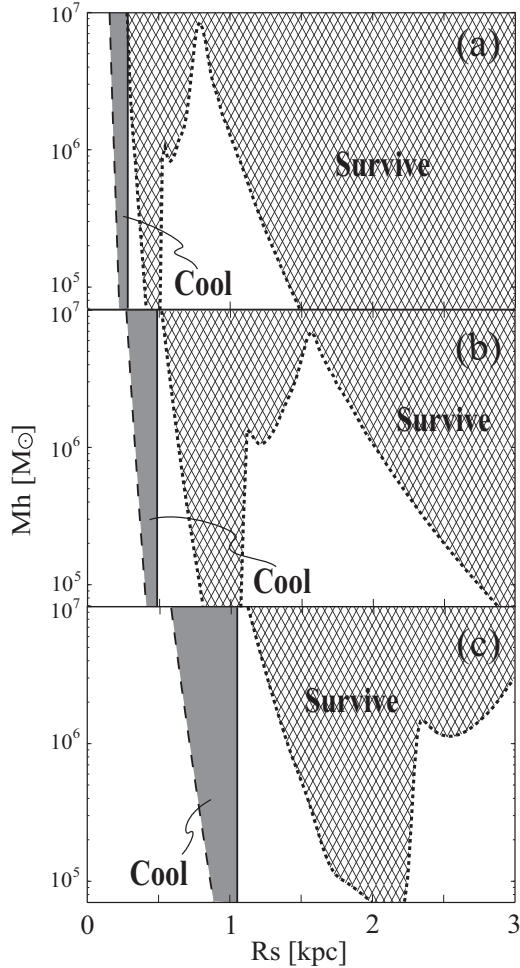}  
  \caption{Regions in which the shocked gas is cooled and the gas in
  the dark halos is not evacuated on the $R_{\rm S}-M_{\rm dh}$ plane.
  Three panels denoted as (a), (b), (c) represent 
  $E_{\rm SN} = 10^{51}, 10^{52}$, and $10^{53} \rm{erg}$, respectively.
  In the hatched area labeled as ``Survive'' ($v_{\rm tot} < V_{\rm esc}$), 
  the shocked gas is bounded in the halo potential. On the other hand,
  in the shaded region denoted as ``Cool'', postshock gas cools
  below $T_{\rm vir}$.}
  \label{fig:fig5}
 \end{center}
\end{figure}

Now, we are ready to combine all conditions to find the criteria for
positive/negative feedback effects by first SNe on the neighboring
dark halos.  Three panels in Figure \ref{fig:fig5} correspond to the
results with $E_{\rm{SN}} = 10^{51},10^{52}$, and $10^{53}\rm{erg}$,
respectively.  Extremely energetic cases with $10^{52}$ and 
$10^{53}\rm{erg}$ can be interpreted as the hypernova or the
pair-instability SN.  
The hatched area bounded by dotted curves (labeled as ``Survive'')
denotes the regions in which $v_{\rm tot} < V_{\rm esc}$ is satisfied.
The shaded region denoted as ``Cool'' represents the condition in which
the cooling rate is high enough for the gas in the halo to proceed
further gravitational contraction.  

First of all, gas components in dark matter halos considered here 
($M \leqslant 10^7\rm{M_ \odot}$) are blown away by the shock momentum
if the halos are close enough to the SN center.  The critical distance
within which the gas is evacuated falls within the range 
$\sim 0.3-1.5$kpc depending on the SN energy and mass of the halo. 
In case we assume normal core-collapse SN ($E_{\rm SN}=10^{51}$erg),
the critical distance is $\sim 0.3-0.5$ kpc, whereas it is $\sim 1-2$kpc
for pair-instability SN ($E_{\rm SN}=10^{53}$erg). The mass dependence
is not so strong, but basically the low-mass halos are more fragile than
the massive halos.  In addition, we also note that the boundaries of the
'Survive' regions have complicated structures at $R_{\rm S}\ga 1$kpc,
reflecting the shape of the cooling function.  Recent studies suggested
that halos further than $\sim 1.5\rm{kpc}$ from the original star will not be
reached by the blast.  In the present paper, we include distances
greater than this in our analysis, but we are doing so only for
completeness. 

The region where ``Survive'' and ``Cool'' are compatible with each other
correspond to the case that the shock-heated gas successfully cooled and
collapse in the dark halo. However, it is clear that the halos which can
survive the evacuation by the SN shock and collapse do not exist.
Consequently, the SN feedback has basically negative effects on the star
formation in surrounding halos. 

\section{DISCUSSION}
The nearby dark halos come under the influence of the radiation from the 
SN progenitor star on the satellite halos prior to the explosion.  Thus,
many low-mass dark halos might be photoevaporated by the UV flux of
progenitor star.  One-dimensional radiation hydrodynamics simulations
are performed by \cite{AS07} on this issue.  They found that effects of
radiative feedback on the gravitational contraction of the gas in
low-mass halos ($M_{\rm dh} \la 10^{6}\rm{M_\odot}$) are very
complicated. The feedback effects are qualitatively different depending
on the distance from the source star, evolutionary stage of the halo at
the onset of radiative feedback, and mass of the halo.  
\citet{Wha08a} also have investigated on this problem by two-dimensional
radiation hydrodynamics simulations, for a single halo with mass of 
$M_{\rm dh} = 1.35\times 10^{5}\rm{M_\odot}$, in which they take into
account a detailed evolutionary stage of the halo.  They found that most
of the gas in this halo is photoevaporated by the SN progenitor
star prior to its death at very early evolutionary stages of the
halo. On the other hand, they indicated that the {\it I}-front could not
reach the core of the halo and this core survived quite well if central
densities of the halo rose beyond $50\rm{cm^{-3}}$.   
\citet{Yoshida07} also performed three-dimensional radiation
hydrodynamics simulations with realistic cosmological density field.
They found the H {\scriptsize II} region of
100$\rm{M_\odot}$ POPIII star extend to $\sim 1$kpc, within which most
of low-mass halos are photoevaporated.   
However, the number of survived halos through UV flux from progenitor star
is still under debate, since we only have the results by multidimensional
simulations with restricted parameter space (mass of the halo, mass of
the progenitor star).  In any case, it is worth to investigate the
effects of SN shock on the nearby halos with various parameters by
analytical calculations. 

In our study, we completely ignore mass loss in the neighbor halos due
to photoevaporation by the progenitor star.  Therefore, it must be noted
that our results are taken to be a lower limit on the damage done by the
expanding remnant to the halo because the halos the SN blast actually
encounters will have less gas.  Also the negative feedback effect we
have found applies only to the one scenario, in which the SNR
increasingly sweeps up mass and then collides with the neighboring
halo. It could be that SN directly form more stars by other means than
those they quench in nearby halos in our mechanism.

\section{CONCLUSION}
We have studied the feedback effects by first SNe with 
$E_{\rm{SN}} = 10^{51}, 10^{52}$, and $10^{53}\rm{erg}$ on their
neighboring dark matter halos. Consequently, the conclusion can be
summarized as follows.  We find that the star formation activities in
the neighboring dark matter halos ($M\leqslant 10^7\rm{M_\odot}$) are
basically suppressed in case they are located close to the SN center,
because of the gas evacuation effect.  The critical distance within
which the gas is blown away falls within the range $\sim 0.3-1.5$kpc 
depending on the SN energy and the halo mass.  In case we assume normal
core-collapse SN ($E_{\rm SN}=10^{51}$erg), the critical distance is
$\sim 0.3-0.5$ kpc, whereas it is $\sim 1-1.5$kpc for pair-instability
SN ($E_{\rm SN}=10^{53}$erg).  In addition, we find there is very little
window in the parameter space where star formation activities in dark
halos are induced or promoted by neighboring SN. 

\acknowledgements
We thank the anonymous referee for critical comments to improve the
paper.  We also thank T. Kitayama, D. Sato, M. Umemura, \&
K. Ohsuga for fruitful discussions and useful comments.  The analysis
has been made with computational facilities at Center for Computational
Sciences in University of Tsukuba and Rikkyo University. 
This work was supported in part by Ministry of Education, Culture,
Sports, Science, and Technology (MEXT), Grants-in-Aid, Specially
Promoted Research 16002003.

\end{document}